\documentclass[pra ,showpacs,showkeys,twocolumn]{revtex4-1}

\usepackage{graphicx}
\usepackage{latexsym}
\usepackage{amsmath}
\usepackage{amssymb}
\usepackage{amsfonts}
\usepackage{color}
\usepackage{cases}
\usepackage[normalem]{ulem}

\begin{document}


\title{Noise-induced transition from superfluid to vortex state in two-dimensional nonequilibrium polariton condensates}

\author{Vladimir N. Gladilin}

\author{Michiel Wouters}

\affiliation{TQC, Universiteit Antwerpen, Universiteitsplein 1,
B-2610 Antwerpen, Belgium}

\date{\today}

\begin{abstract}
We study  the Berezinskii-Kosterlitz-Thouless mechanism for
vortex-antivortex pair formation in two-dimensional superfluids for
nonequilibrium condensates. Our numerical study is based on a
classical field model for driven-dissipative quantum fluids that is
applicable to polariton condensates. We investigate the critical
noise needed to create vortex-antivortex pairs in the systems,
starting from a state with uniform phase. The dependence of the
critical noise on the nonequilibrium and energy relaxation
parameters is analyzed in detail.
\end{abstract}


\maketitle

\section{Introduction}

With the advent of synthetic quantum systems, the interest in
driven-dissipative many-body systems has grown substantially in the
last decade. Where particles in ultracold atomic gases can to a very
good approximation be conserved, losses can be engineered
\cite{ott16} or are unavoidable in strongly interacting Bose gases
\cite{bose-unit}. In systems based on electromagnetic degrees of
freedom, both in the microwave \cite{fitzpatrick17} and optical
domain \cite{carusotto-ciuti,brenecke13}, cavity losses are often
not negligible. In order to reach a steady state, some driving of
the system is then necessary to compensate for the losses. This
raises the question on the modifications of the steady state with
respect to the thermal equilibrium state in conservative systems.

Here, we will consider the case of two-dimensional weakly
interacting bosons that are subject to single particle losses, which
are compensated by a nonresonant drive. These systems are realized
by microcavity polariton condensates \cite{Kasprzak06}, but there
may be also the possibility to construct them with ultracold atoms
\cite{ott16}. Microcavity polaritons are hybrid light-matter
quasiparticles resulting from the strong coupling between an
excitonic transition in a quantum well and a photonic mode in an
enveloping microcavity. From their photonic component, they inherit
a light effective mass, enhancing the spatial coherence, whereas
from their excitonic component, they inherit interactions (see Refs.
\cite{rodriguez17,delteil19} for the experimental determination of
the interaction constant). Under nonresonant excitation, a large
density of excitons is created, which subsequently relax to the
lower polariton region. At equilibrium, the two-dimensional bose gas
features a Berezinskii-Kosterlitz-Thouless (BKT) phase transition:
for increasing temperature,  thermally excited vortex-antivortex
pairs become unbound, resulting in the loss of superfluidity. A
natural question is then how this transition will be affected by
losses and driving.

From the experimental side, this physics was addressed in Ref.
\cite{caputo2016}. In their system with a long polariton life time,
they did find a phase transition that was interpreted as the binding
to unbinding transition of vortex-antivortex pairs, reminiscent of
the equilibrium situation.

Already at the level of small phase fluctuations, nonequilibrium
systems behave differently from their equilibrium counterparts.
Where for the latter, the phase dynamics is for small fluctuations
to a good approximation described by a linear equation, in the
nonequilibrium case, a nonlinear term appears, which brings them in
the Kardar-Parisi-Zhang universality class
\cite{wachtel,sieberer,he15,altman15,sieberer16,squizzato18,gladilin14}.

Some first theoretical insight in the modification of the BKT
transition due to the nonequilibrium condition can be gained by
considering the modification of vortices when going away from
equilibrium. It turns out that the gain/losses introduce an
additional current with the vortex core as its source. The phase
profile is consequently deformed, resulting in a spiral wave
\cite{aranson2002}. This modification in the vortex flow field
subsequently affects the interactions between vortices and
antivortices: when the outward flows are more important than the
usual azimuthal flows, the interaction between vortex and antivortex
becomes repulsive. These repulsive interactions hamper the
vortex-antivortex recombination, enabling the formation of
vortex-antivortex clusters with a very long life time
\cite{gladilin18}. A renormalization group based approach has shown
that these repulsions are fatal for the superfluid phase. The
renormalization flow always goes toward the normal phase, even
though this physics may manifest itself only at very large distances
\cite{wachtel}.

In order to shed further light on the phase diagram of 2D
nonequilibrium polaritons, we  resort here to numerical simulations
of the noisy generalized Gross-Pitaevskii equation (ngGPE). This
equation can be derived within the truncated Wigner approximation
\cite{wouters09} or with the Keldysh field theory formalism
\cite{szymanska07, sieberer14} and has been widely used to model
nonresonantly pumped polariton condensates \cite{16, kalinin18,
comaron18, bobrovska17, estrecho18,ohadi18}. At equilibrium, when
the Bose gas is fully characterized by its density, temperature and
interaction constant, the critical temperature is found to equal
$T_c \approx 2\pi \hbar^2 n/[m \ln(380 \hbar^2/mg)]$, where $n$ is
the density, $m$ the mass and $g$ the interaction strength
\cite{prokofev01}. Away from equilibrium however, there are more
microscopic parameters that enter the theoretical description. We
investigate here how they affect the critical noise strength (the
analog of the temperature out of equilibrium).

In a previous theoretical study~\cite{gladilin18}, based on  a noise
free generalized GPE, we found that starting from an initial state
with a large number of vortices, several can survive in the steady
state, because of the repulsive interactions between vortices and
antivortices. We even found that these can form quite regular
structures. With this physics understood, we will start here from
the opposite initial condition with a homogeneous phase. In
polariton condensates, such an initial condition can be achieved by
sending a resonant pulse with a flat phase profile. As expected, we
find that only when sufficiently strong noise is present,
vortex-antivortex pairs can be formed in the subsequent time
evolution. We will show that the pair production shows a well
defined noise threshold, allowing us to draw a phase diagram for the
system.

For systems that are far from equilibrium, we have shown
\cite{30,gladilin18} that the gGPE predicts a self-acceleration of
vortices and production of new pairs in vortex-antivortex
collisions, leading to chaotic dynamics. In this parameter regime,
we find that a moderate noise suppresses this mechanism, leading to
a stabilization of the system.

The structure of the paper is as follows. In Sec. \ref{sec:model},
our model for nonequilibrium condensation is recapitulated. The
phase diagram is discussed in Sec. \ref{sec:results} and conclusions
are drawn in Sec. \ref{sec:concl}.

\section{Model \label{sec:model}}

We consider nonresonantly excited two-dimensional polariton
condensates. In the case of sufficiently fast relaxation in the
exciton reservoir, this reservoir can be adiabatically eliminated
and the condensate is described by the noisy generalized
Gross-Pitaevskii equation~\cite{wouters09,szymanska07,
sieberer14,15,16,a31}
\begin{eqnarray}
({\rm i}-\kappa)\hbar \frac{\partial \psi}{\partial t} =&&
\left[-\frac{\hbar^2\nabla^2}{2m} +g |\psi|^2 \right. \nonumber
\\
&&\left.+\frac{{\rm i}}{2} \left(\frac{P}{1+|\psi|^2/n_s}-\gamma
\right) \right] \psi+\sqrt{D}  \xi . \label{ggpe}
\end{eqnarray}
Here $m$ is the effective mass and the contact interaction between
polaritons is characterized by the strength $g$. The imaginary term
in the square brackets on the right hand side describes the
saturable pumping (with strength $P$ and saturation density $n_s$)
that compensates for the losses ($\gamma$). We take into account the
energy relaxation $\kappa$ in the condensate~\cite{38,39}. The
complex stochastic increments have the correlation function $\langle
\xi^*(x,t) \xi(x',t') \rangle=2 \delta({\bf r}-{\bf r}')
\delta(t-t')   $.

For polariton condensates, the validity of Eq. \eqref{ggpe} is not
always straightfoward to justify. The repulsive interactions between
the condensate and the exciton reservoir may lead to an effective
attraction between the polaritons, leading to instability for a
positive polariton mass \cite{baboux18,bobrovska14,bobrovska17}.
This unstable state can be stabilized by a negative effective mass,
that can be obtained in a polariton microcavity lattice
\cite{baboux18}. We will not consider the instability physics in
this work and assume positive mass and positive interactions (but
the physics remains unaltered when the signs of the mass,
interaction strength and energy relaxation parameter ($\kappa$) are
simultaneously changed).

It is then convenient to rewrite Eq.~(\ref{ggpe}) in a dimensionless
form, by expressing the particle density $|\psi|^2$ in units of
$n_0\equiv n_s(P/\gamma-1 )$, time in units of $\hbar/(gn_0)$,
length in units of $\hbar/\sqrt{2mgn_0}$, and noise intensity in
units of $\hbar^3 n_0/(2m)$:
\begin{eqnarray}
({\rm i}-\kappa)\frac{\partial\psi}{\partial t}=&& \left[-\nabla^2
+|\psi|^2  \phantom{\frac{\psi^2}{\psi^2}}\right. \nonumber
\\
&&\left.+{\rm i}c\frac{1-|\psi|^2}{1+\nu |\psi|^2} \right] \psi
+\sqrt{D} \xi . \label{ggpe2}
\end{eqnarray}
Besides the noise intensity $D$, equation~(\ref{ggpe2}) contains
three other dimensionless scalar parameters: $\kappa$ characterizes,
as described above, damping in the system dynamics,
$c=\gamma/(2gn_s)$ is a measure of the deviation from equilibrium,
and $\nu=n_0/n_s$ is proportional to the relative excess of the
 pumping intensity $P$ over the threshold intensity. In the
absence of noise, Eq. \eqref{ggpe2} has the steady state solution
$\psi = \sqrt{n} e^{- i n t}$, where the density $n$ satisfies
\begin{equation}
\kappa n = \frac{c(1-n)}{1+\nu n}, \label{eq:dens_eq}
\end{equation}
so that $n$ is a decreasing function of $\kappa$.

For weak noise, the fluctuations on top of this homogeneous solution
can in first approximation be analyzed within the Bogoliubov
approach \cite{chiocchetta13}, but a more refined analysis has
revealed that the dynamics of the phase fluctuations is in the KPZ
universality class \cite{gladilin14, ji15, altman15,
he15,squizzato18,he17}, where the phase nonlinearity cannot be
neglected.

Here, we will study the critical noise needed to create vortices. In
order to address this problem, Eq. (\ref{ggpe2}) is solved
numerically using the same finite-difference scheme as in
Ref.~\cite{30}. Specifically, we use periodic boundary conditions
for a square of size $L_x=L_y=40$ with grid step equal to 0.2
corresponding to a cutoff in momentum space equal to $K_c=5 \pi$.
The model \eqref{ggpe} is a classical field model, that suffers from
the usual ultraviolet catastrophe \cite{prokofev01}. This implies
that our results will be cutoff dependent. Below, we will discuss
this dependence and how it will affect the comparison of our
theoretical results with experiments.

\begin{figure} \centering
\includegraphics*[width=0.9\linewidth]{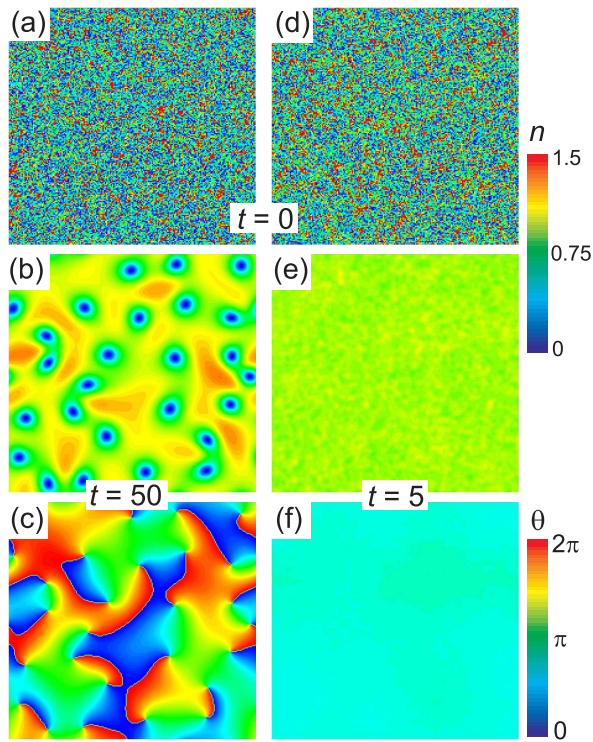}
\caption{Distributions of the particle density $n$ for $c=1.2$,
$\nu=1$, $\kappa=0.01$, $D=0.015$ and $t_D=100$ at the time moment
$t=0$ when the noise is switched off (a) and at $t=50$ (b).
Distribution of the phase $\theta$ of the order parameter at $t=5$
is shown in panel (c). Panels (d), (e) and (f): same as in panels
(a), (b) and (c), respectively, but for $D=0.014$, $t_D=600$ and a
shorter time delay after switching off the noise ($t=5$).
\label{example}}
\end{figure}

When analyzing the noise-induced BKT transition, each run starts
from a uniform condensate distribution at $D=0$. Then we apply noise
with a fixed nonzero intensity $D$ during a time interval $t_D$. The
detection of vortex pairs in the presence of a strong noise is
somewhat tricky, but fortunately in the absence of noise their
annihilation time is known to be rather long even at very weak
nonequilibrium~\cite{kulczy,Comaron17} and this time strongly
increases with increasing $c$~\cite{gladilin18}. For these reasons,
it is more convenient to check the presence of vortex pairs sometime
later after switching off the noise. The used time delay (typically
few our units of time) is sufficient for significant relaxation of
the noisy component in the density and phase distributions of the
condensate and, at the same time, is too short for vortex pair
annihilation. To determine the critical noise for the BKT
transition, $D_{\rm BKT}$, we use the following criterion. If for a
noise intensity $D$ vortex pairs are present after a noise exposure
time $t_D$ (and hence $D>D_{\rm BKT}$), while for a certain noise
intensity $D^\prime<D$ no vortex pairs appear even at noise
exposures few times longer then $t_D$, then $D^\prime$ lies either
below $D_{\rm BKT}$ or above $D_{\rm BKT}$ and closer to $D_{\rm
BKT}$ then to $D$. Therefore, the critical noise intensity can be
estimated as $D_{\rm BKT}=D^\prime \pm (D-D^\prime)$. An
illustration is displayed in Fig.~\ref{example}. For $D=0.015$ and
noise exposure time $t_D=100$, a noisy density distribution shown in
Fig.~\ref{example}(a) evolves after switching off the noise into a
pattern with clearly seen vortices and antivortices, which persist
during a relatively long time [see Figs.~\ref{example}(b) and (c)].
For a slightly lower noise intensity, $D=0.014$, and significantly
longer noise exposure time, $t_D=600$ , after switching off the
noise the corresponding noisy distribution [Fig.~\ref{example}(d)]
rapidly relaxes towards a uniform vortex-free state
[Figs.~\ref{example}(e) and (f)]. According to our approach, the
estimate for the critical noise in this case is $D_{\rm
BKT}=0.014\pm 0.001$.


\section{Results and Discussion \label{sec:results} }

\begin{figure} \centering
\includegraphics*[width=0.9\linewidth]{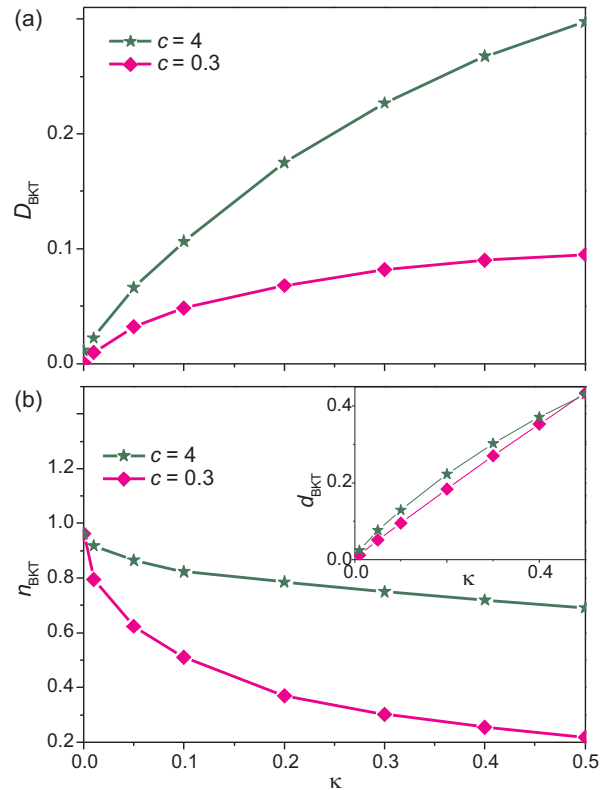}
\caption{Noise intensity $D_{\rm BKT}$ corresponding to the BKT
transition [panel (a)] and average density of the condensate $n_{\rm
BKT}$ at $D=D_{\rm BKT}$ [panel (b)] as a function of damping at
$\nu=1$ and two different values of the nonequilibrium parameter
$c$. The error bars are smaller than the symbol size. Inset of panel
(b): ratio $d_{\rm BKT}=D_{\rm BKT}/n_{\rm BKT}$ for $c=4$ (stars)
and $c=0.3$ (diamonds). \label{dkap}}
\end{figure}

In Fig.~\ref{dkap}(a), the critical noise intensity corresponding to
the BKT transition, $D_{\rm BKT}$, is shown as a function of the
damping parameter $\kappa$ for the cases of moderate ($c=0.3$) and
strong ($c=4$) deviations from equilibrium. In both cases the
critical noise $D_{\rm BKT}$ is seen to considerably increase with
$\kappa$, despite the fact that the average density of the
condensate at $D=D_{\rm BKT}$ is a monotonously decreasing function
of $\kappa$ [see also  Eq. \eqref{eq:dens_eq}], this decrease being
especially pronounced at smaller $c$ [see Fig.~\ref{dkap}(b)].

The inset of Fig.~\ref{dkap}(b) shows the ratio $d_{\rm BKT}\equiv
D_{\rm BKT}/n_{\rm BKT}$ as a function of $\kappa$. For the two very
different values of the nonequilibrium parameter $c$, the curves
$d_{\rm BKT}(\kappa)$ appear to lie relatively close to each other.
One can also notice that at $c=0.3$ the dependence of $d_{\rm BKT}$
on $\kappa$ is nearly linear (except for the smallest values of
$\kappa$).

At $\kappa \gg 1$ linear scaling of the noise intensity with
$\kappa$ directly follows from an additional rescaling of time
[$t\rightarrow t / \kappa$; see Eq.~(\ref{ggpe2})]. Our numerical
results show that this linear dependence of $d_{\rm BKT}$ on
$\kappa$ remains to good approximation unaltered even when $\kappa
<1$.

At equilibrium ($c=0$), the increase of the critical noise with
increasing $\kappa$ can be understood by making the connection with
the thermal equilibrium case.  For vanishing nonequilibrium and
$\kappa \gg 1$, our equation reduces to model A dynamics
\cite{hohenberg-halperin}, which has a Boltzmann-Gibbs steady state
distribution at temperature  $T = D/2\kappa$. In this limit, the
transition occurs at the equilibrium BKT temperature
\cite{prokofev01}, where $T_{\rm BKT} = \eta n$, with $\eta$ a
numerical cutoff dependent constant. For the critical noise, one
then obtains $D_{\rm BKT} =2 \eta \kappa n$.

Numerically, we have observed that for $c\rightarrow 0$, the
critical noise obeys this relation, even for $\kappa < 1$. The fact
that the dissipative part of the dynamics does not alter the steady
state can be understood from the following argument. It is well
established that pure Gross-Pitaevskii dynamics ($D=\kappa=c=0$)
samples the thermal equilibrium state in the microcanonical
ensemble, at an energy determined by the initial state. Similarly,
the Langevin dynamics [when omitting the $i$ in the l.h.s. of Eq.
\eqref{ggpe2} and taking $c=0$] samples the phase space according to
the canonical ensemble at a temperature determined by the balance
between noise and dissipation. Since the thermal state is the steady
state of both the GP and Langevin dynamics, it is natural that the
steady state of the combined dynamics is also at thermal
equilibrium, with the temperature determined by the Langevin part.

\begin{figure} \centering
\includegraphics*[width=0.9\linewidth]{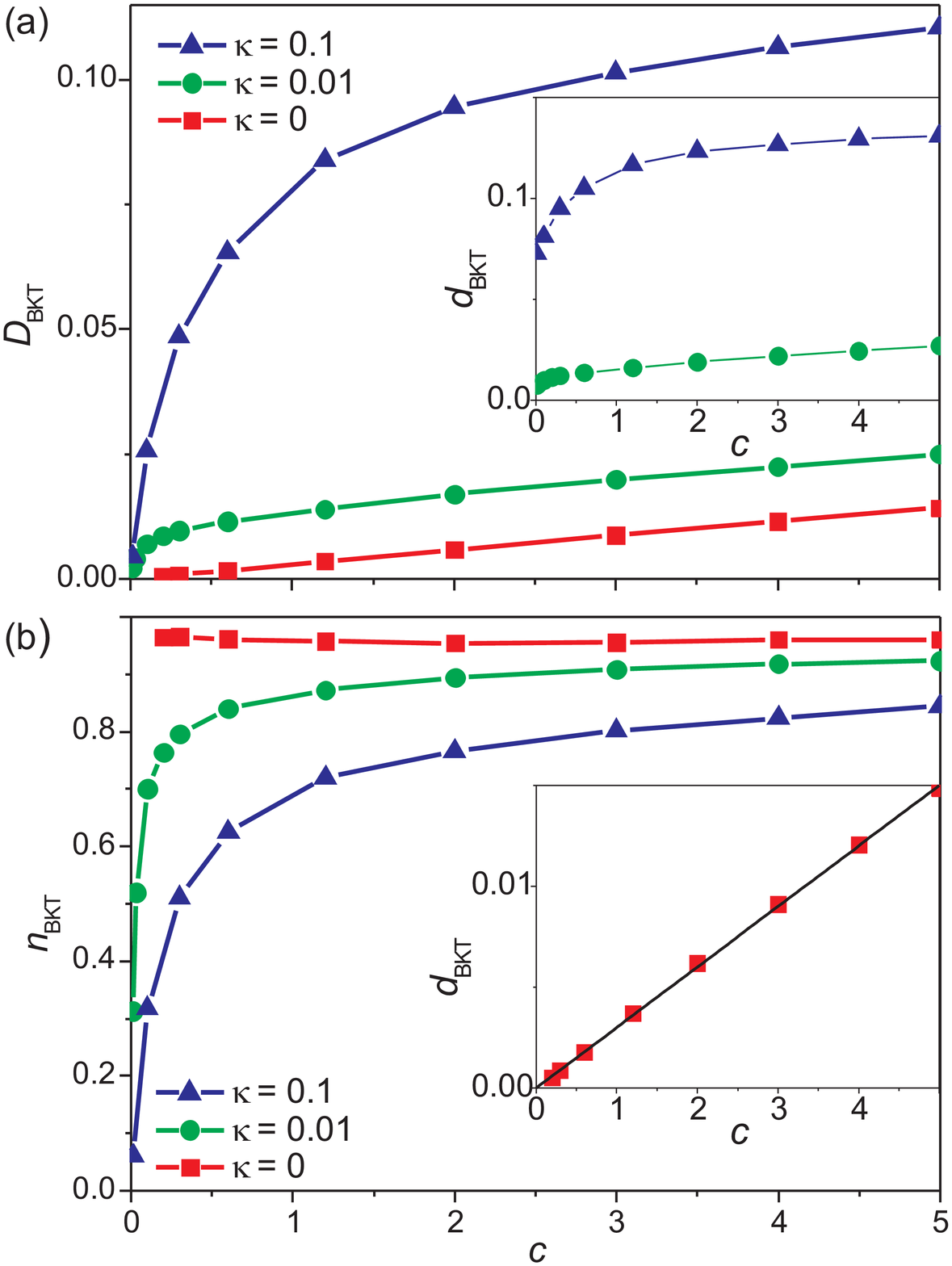}
\caption{Noise intensity $D_{\rm BKT}$ corresponding to the BKT
transition [panel (a)] and average density of the condensate $n_{\rm
BKT}$ at $D=D_{\rm BKT}$ [panel (b)] as a function of the
nonequilibrium parameter $c$ at $\nu=1$ and different values of the
damping parameter $\kappa$. The error bars are smaller than the
symbol size. Inset of panel (a): ratio $d_{\rm BKT}=D_{\rm
BKT}/n_{\rm BKT}$ for $\kappa=0.1$ (triangles) and $\kappa=0.01$
(circles). Inset of panel (b): ratio $d_{\rm BKT}=D_{\rm BKT}/n_{\rm
BKT}$ for $\kappa=0$ (squares). The black line corresponds to the
linear fitting $d_{\rm BKT}=a c$, with $a=0.003$.\label{dc}}
\end{figure}

As implied by the results displayed in Fig.~\ref{dkap}(a), the
critical noise intensity $D_{\rm BKT}$ increases when moving away
from equilibrium. The dependence of $D_{\rm BKT}$ on the
nonequlibrium parameter $c$ is further illustrated in
Fig.~\ref{dc}(a) for the cases of  weak ($\kappa=0.01$ and 0.1) and
zero damping. At nonzero damping the increase of $D_{\rm BKT}$ with
$c$ is partly related to the simultaneous increase of the average
condensate density $n_{\rm BKT}(c)$ shown in Fig.~\ref{dc}(b). The
latter originates from the growing contribution of the pumping-loss
term in Eq.~(\ref{ggpe2}) [the last term in the square brackets in
Eq. \eqref{ggpe2}]. Indeed, this term tends to keep the condensate
density as close to 1 as possible. As a result, the degrading effect
of damping or noise on the condensate density weakens with
increasing $c$. This can be seen, e.g., from Fig.~\ref{dkap}(b) by
comparing to each other the values of $n_{\rm BKT}$ at different
$c$: for each fixed, not too small $\kappa$, the value of $n_{\rm
BKT}|_{c=4}$ is significantly larger than $n_{\rm BKT}|_{c=0.3}$,
even though the density $n_{\rm BKT}|_{c=4}$ corresponds to a
considerably higher noise intensity than that for $n_{\rm
BKT}|_{c=0.3}$ [see Fig.~\ref{dkap}(a)].

At the same time, as follows from the behavior of the ratio $d_{\rm
BKT}\equiv D_{\rm BKT}/n_{\rm BKT}$ [see the inset of
Fig.~\ref{dc}(a)], the increase of $D_{\rm BKT}$ with $c$ is
considerably faster as compared to that of $n_{\rm BKT}$. This means
that the effect of pumping-loss processes on $D_{\rm BKT}$ cannot be
explained solely by the corresponding stabilization of the average
condensate density against noise and damping. This becomes even more
evident when looking at the results for $\kappa=0$. Indeed, while
the critical noise $D_{\rm BKT}$ demonstrates a clear rise with
increasing $c$ [see Fig.~\ref{dc}(a)], the average density $n_{\rm
BKT}(c)$ remains almost constant [see Fig.~\ref{dc}(b)]. At small
$c$, the values of $n_{\rm BKT}$ even slightly decrease with $c$
suggesting that the stabilizing effects of the pumping-loss
processes do not completely compensate the average-density reduction
caused by the increase of the noise intensity at the BKT transition.
However, an important aspect of the the pumping-loss processes,
which is not directly reflected in the {\it average} density, is
that they impede formation of deep {\it local} suppressions of the
condensate density. Those deep density suppressions, together with
the appropriate phase gradient, are necessary prerequisites for
vortex pair formation and hence for the BKT transition. The
discussed numerical results imply that just the stabilizing effect
of the pumping-loss term on the local density of the condensate
governs the increase of $D_{\rm BKT}$ with $c$ at $\kappa
\rightarrow 0$. At $\kappa = 0$ the dependence of $D_{\rm BKT}$ on
$c$ becomes close to linear [see Fig.~\ref{dc}(a)], while for
$d_{\rm BKT}(c)$ a linear function $d=ac$ with $a=0.003$ provides a
nearly perfect approximation [see Fig.~\ref{dc}(b)].

This interpretation of our numerical results can be elucidated by a
linear analysis of the fluctuations. Writing the field as
$\psi(x,t)=\sqrt{1+\delta n(x,t)}e^{i \theta(x,t)}$, one obtains in
linearized approximation for the Fourier components
\begin{eqnarray}
\frac{\partial}{\partial t} \left(
\begin{array}{c}
\theta_k \\
n_k
\end{array}
\right) =&& \left(
\begin{array}{cc}
-\kappa \epsilon_k &  -\frac{\epsilon_k}{2}-1 \\
2 \epsilon_k       &  - \frac{2c}{1+\nu}+4 \kappa
\end{array}
\right) \left(
\begin{array}{c}
\theta_k \\
n_k
\end{array}
\right) \nonumber \\ &&+\left(
\begin{array}{c}
\sqrt{D} \xi^{(\theta)}_k \\
\sqrt{4 D} \xi^{(n)}_k
\end{array}
\right),
\end{eqnarray}
where $\epsilon_k=k^2$. The white noise in Fourrier space has the
correlation function $\langle \xi^{(\theta)}(k,t)
\xi^{(\theta)}(k',t) \rangle = 4\pi \delta(k+k') \delta(t-t') $,
analogous for $\xi^{(n)}$, and $\langle \xi^{(\theta)} \xi^{(n)}
\rangle=0$.

The steady state density fluctuations can be obtained in closed
form. In the limit of large $k$, they simplify to

\begin{numcases}{\langle n_k n_{k'} \rangle = }
   2 \pi \delta(k+k')
\frac{4 D (2+\kappa^2)}{\kappa \epsilon_k} & $\kappa \neq 0$
   \\
   2 \pi \delta(k+k') \frac{8 D  (1+\nu)}{c}  & $\kappa = 0$
\end{numcases}
For the density fluctuations in real space, we then obtain for a
sufficiently large momentum cutoff $K_c$
\begin{numcases}{\langle \delta n^2(x) \rangle \propto }
    \frac{D}{\kappa}(2+\kappa^2) \ln K_c & $\kappa \neq 0$ \label{eq:n2knz}
   \\
    \frac{D(1+\nu)}{c} K_c^2  & $\kappa = 0$ \label{eq:n2kz}
\end{numcases}
The local density fluctuations obtained here in the linear
approximation show behavior that is in line with what was observed
numerically for the BKT transition. Density fluctuations are first
suppressed by the damping $\kappa$ and for $\kappa=0$ by the
nonequilibrium parameter $c$.

Note the very different dependence of both results on the momentum
cutoff. In the presence of a nonzero damping, the cutoff dependence
is a very weak logarithm, where in its absence, it becomes
quadratic. This difference is due to the fact that the momentum
space density tends to a constant in the $\kappa=0$ case, where it
decays as $k^{-2}$ for $\kappa \neq 0$. Even the latter is not fast
enough to ensure convergence in two dimensions, hence the
logarithmic divergence of the density fluctuations.

The values of $D_{\rm BKT}$, obtained from the numerical
simulations, correlate with the above analysis. At $c=1.2$, $\nu=1$,
and $\kappa=0$, an increase of the grid step from 0.2 to 0.4 leads
to a change of the numerically determined $D_{\rm BKT}$ from
$0.00355\pm 0.0001$ to $0.0139 \pm 0.0001$ that is in line with the
quadratic dependence of $\langle n^2(x) \rangle$ on $K_c$ in the
absence of energy relaxation [see Eq.~\eqref{eq:n2kz}]. At the same
time, for nonzero damping ($\kappa \geq 0.01$) the relative
difference between the values of $D_{\rm BKT}$ corresponding to the
grid step 0.4 and 0.2 is about 40\% for $\kappa=0.01$ and does not
exceed 15\% for $\kappa \geq 0.1$. The critical density increases
somewhat when increasing the grid step, such that $d_{\rm BKT}=
D_{\rm BKT}/n_{\rm BKT}$ increases only by about 10\% for $\kappa
\geq 0.1$. Our analytical arguments and those of Ref.
\cite{prokofev01} suggest a logarithmic dependence on the cutoff,
but we were not able to test this scaling numerically due to a lack
of accessible range in the grid spacing. {Choosing the grid spacing
much larger than 0.4, it becomes too coarse to give a good
description of the vortex cores, while choosing a grid spacing
smaller than 0.2 slows down the calculations too much (the maximal
kinetic energy increases quadratically with momentum cutoff).
Moreover, decreasing the grid spacing below 0.2 would give a kinetic
energy that is typically larger than what is required to justify the
lower polariton approximation. }

As seen from Eq.~(\ref{ggpe2}), the pumping-loss term increases in
magnitude when decreasing the parameter $\nu$. One can expect,
therefore, that a decrease of $\nu$ leads to an increase of the
critical noise $D_{\rm BKT}$. Our simulations confirm this
expectation but, at the same time, show that the influence of $\nu$
on $D_{\rm BKT}$ is relatively weak for nonvanishing damping. At
$\kappa=0.01$ a decrease of $\nu$ by one order of magnitude (from 1
down to $0.1$) results in an increase of $D_{\rm BKT}$ approximately
by 4\% at $c=0.3$ and by 12\% at $c=4$. The corresponding increase
of $d_{\rm BKT}$ is about 20\% for both $c=0.3$ and 4. In other
words, the effect of $\nu$ on the critical noise is rather minor as
compared to the much stronger impact of $\kappa$ and $c$.

\begin{figure} \centering
\includegraphics*[width=0.9\linewidth]{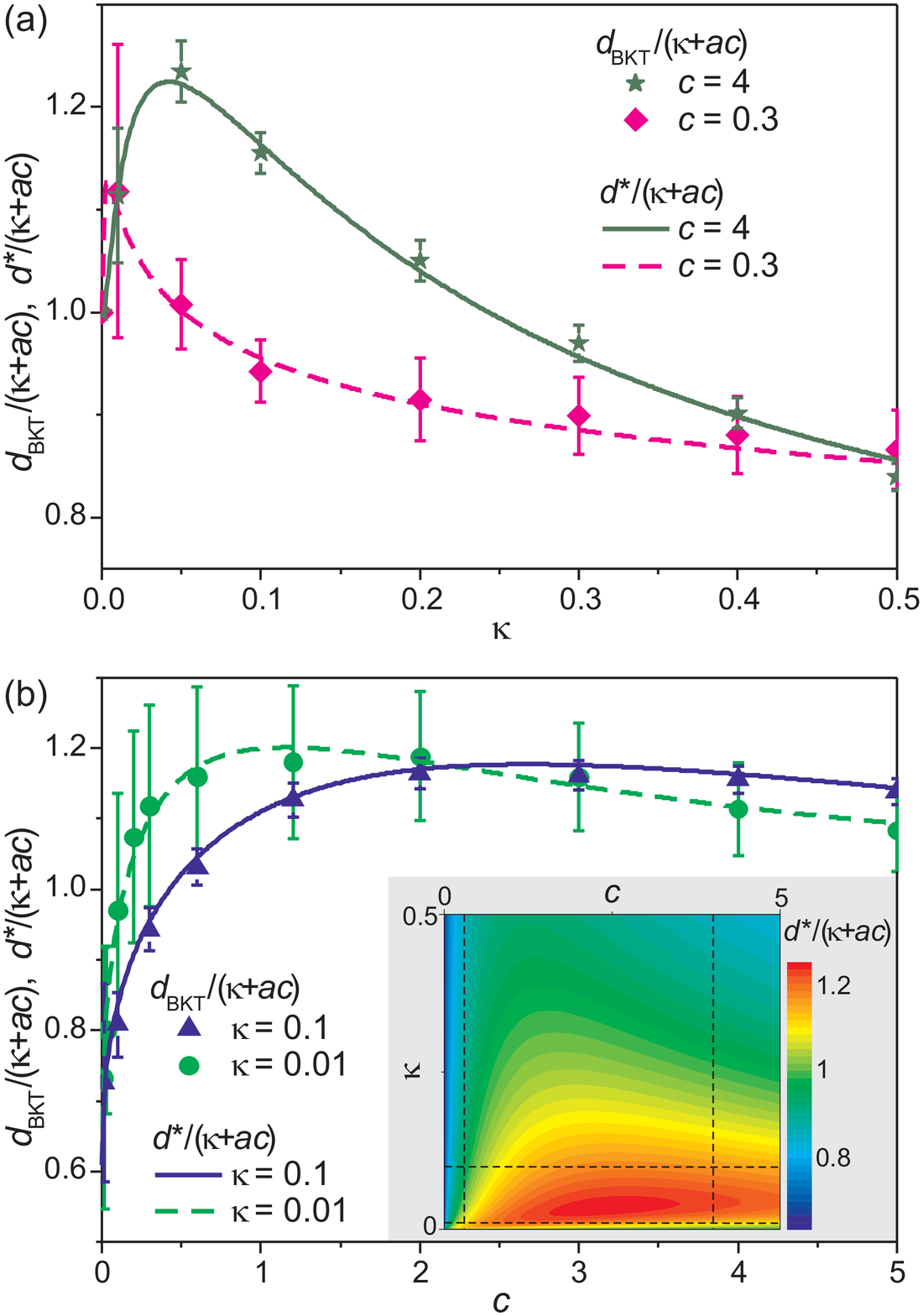}
\caption{Renormalized noise intensity $d_{\rm BKT}/(\kappa+ac)$
(symbols) and its fitting by $d^*/(\kappa+ac)$ (lines) as a function
of the damping parameter $\kappa$ [panel (a)] and nonequilibrium
parameter $c$ [panel (b)].The inset shows the dependence of
$d^*/(\kappa+ac)$ on both $c$ and $\kappa$. Straight dashed lines
correspond to the parameter values covered by the curves in panels
(a) and (b). \label{dnorm}}
\end{figure}

Let us now look in some more detail at the simultaneous dependence
of $d_{\rm BKT}$ on $\kappa$ and $c$. Taking into account, on the
one hand, the nearly linear dependence $d_{\rm BKT}$ on $\kappa$ at
large $\kappa$ [see the inset in Fig.~\ref{dkap}(b)] and, on the
other hand, the linear dependence on $c$, $d_{\rm BKT}\approx ac$ at
$\kappa \rightarrow 0$, it seems {reasonable} to consider the
renormalized quantity $d_{\rm BKT}/(\kappa+ac)$. As seen from
Fig.~\ref{dnorm}, all the results of our simulations lie in a rather
narrow interval around 1: $0.6<d_{\rm BKT}/(\kappa+ac)<1.3$. This
suggests that the simple expression $\kappa + 0.003c$ already can
serve as a crude but quite reasonable estimate for $d_{\rm BKT}$ at
any $\kappa$ and $c$. A better approximation of the numerical
results is obtained (see Fig.~\ref{dnorm}) by using the fitting
function, {that summarizes our numerical calculations rather
accurately:}
\begin{eqnarray}
d^*=&& 0.609 \kappa+0.003c \nonumber \\ &&+\frac{ \kappa^{1.411}
c^{0.411} }{2.646\kappa^{0.610}  + 0.0706 c^{0.889} + 1.622
\kappa^{1.514} c^{1.308} }
\end{eqnarray}
Since the numerical results for the dependence of $d_{\rm BKT}$ on
both $\kappa$ and $c$ are seen to be fitted quite well by
$d^*(c,\kappa)$, we may expect that this function can provide also
meaningful results for inter- and extrapolation {[see the inset in
Fig.~\ref{dnorm}(b)]}.

\begin{figure} \centering
\includegraphics*[width=0.9\linewidth]{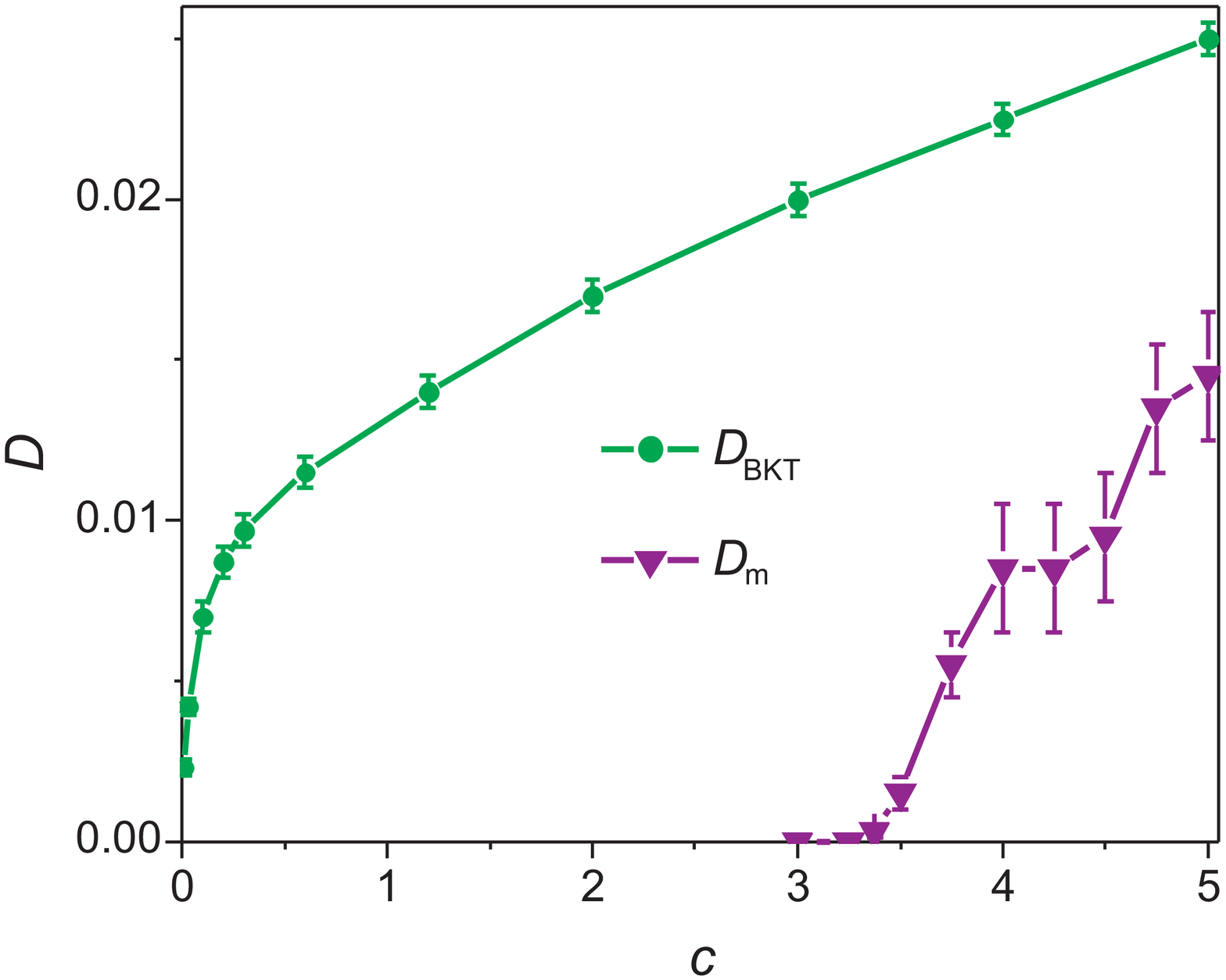}
\caption{Noise intensity $D_{\rm BKT}(c)$ corresponding to the BKT
transition in comparison with $D_{\rm m}(c)$, the maximum noise
intensity, which still allows for vortex pair generation by moving
vortices, at $\nu=1$ and $\kappa=0.01$. \label{dm}}
\end{figure}

Finally, we address the question of whether the BKT transition under
consideration can be influenced by the effect of vortex pair
generation by moving vortices~\cite{gladilin18}. This effect has
been predicted to occur in strongly nonequilibrium condensates when
self-accelerated vortices acquire sufficiently high velocities with
respect to the surrounding condensate. The results
of~\cite{gladilin18} were obtained in the absence of noise. In our
numerics containing the noise term, we have observed  that
fluctuations of the condensate density and currents tend to impede
the acceleration of vortices. Consequently, the generation of vortex
pairs in vortex collisions becomes impossible at high noise
intensities. This is illustrated in Fig.~\ref{dm}, which shows the
upper boundary $D_{\rm m}$ for the noise intensity range, where
generation of new vortex pairs by self-accelerated vortices is
possible, as a function of $c$ at $\nu=1$ and $\kappa=0.01$. The
corresponding calculations are performed, like in
~\cite{gladilin18}, starting with one pinned vortex pair in the
region under consideration and simulating the dynamics during a time
interval $\Delta t \sim 1000$ after depinning of vortices, now in
the presence of noise.

In Fig.~\ref{dm}, the dependence of $D_{\rm m}$ on $c$ is given in
comparison with the critical noise $D_{\rm BKT}$ at the same
$\kappa$ and $\nu$. As seen from Fig.~\ref{dm}, the region of $c$
and $D$, where pair generation by vortices is possible, lies at
 large $c$ ($c > 3.3$) and has no overlap with the curve
$D_{\rm BKT}(c)$. Of course, the latter remains true for any $\kappa
\geq 0.01$. Indeed, the critical noise $D_{\rm BKT}$ increases with
$\kappa$. At the same time, damping slows down vortex motion and
thus impedes pair generation by moving vortices. As a result, the
aforementioned region will shrink with increasing $\kappa$. Our
simulations show that already at $\kappa$ as small as 0.1 the
velocities of vortices are insufficient for pair generation. So, we
can conclude that at $\kappa \geq 0.01$ the BKT transition is not
affected by the processes of pair generation by moving vortices. As
concerns the limit of very weak damping $\kappa< 0.01$, the nearly
linear behavior of $D_{\rm BKT}(c)$, obtained for $\kappa=0$,
manifests no visible peculiarities in the region of large $c$, where
the processes of pair generation by vortices become possible. This
implies that, also in the limit $\kappa \to 0$, the BKT transition
is not considerably affected by these processes.

\section{Conclusions \label{sec:concl}}

We have investigated the critical noise for the spontaneous
formation of unbound vortex-antivortex pairs in a driven-dissipative
bosonic system where particle losses are compensated by nonresonant
pumping. In this work, we have focused on the noise needed to form
vortex-antivortex pairs starting from a uniform phase. In our model,
the critical noise strength depends -- with a suitable choice of
units -- on three dimensionless parameters: the energy relaxation
rate ($\kappa$), the nonequilibrium parameter ($c$) and the gain
saturation parameter ($\nu$).

In the absence of energy relaxation, the nonequilibrium parameter
determines the critical noise strength, but this critical value
depends quadratically on the momentum space cutoff, giving our
numerical results limited predictivity for specific experiments. In
the presence of sufficiently strong energy relaxation, the cutoff
dependence is much weaker and experimentally relevant results can be
extracted from the numerics in this case. In this regime, the
critical noise strength increases both with energy relaxation (as
expected from equilibrium calculations) and with nonequilibrium
parameter $c$. The latter dependence could seem counterintuitive,
because with increasing $c$, vortices and antivortices repel each
other at large distances, which could favor their unbinding. We
interpret the impeding of vortex-antivortex formation further away
from nonequilibrium as a consequence of the reduction of the density
fluctuations. The effect of nonequilibrium  on the
\textit{formation} of vortices is therefore opposite to its effect
on the \textit{annihilation}:  the life time of existing vortices is
dramatically enhanced by nonequilibrium~\cite{gladilin18}.

Quantitatively, we have found that the effect of the nonequilibrium
parameter is actually small for $\kappa \gg 0.003 c$, which is
satisfied when damping is not too small and the system is not too
far from equilibrium. Very far from equilibrium, self acceleration
of vortices can lead to the production of new vortex antivortex
pairs. We have shown here that this pair production ceases for
increasing noise.

Our numerical simulations were performed for systems with
periodic boundary conditions whose size is comparable to
the systems employed in current experiments.
The study of the thermodynamical limit of infinite
system size remains a challenge for both theoretical analysis and
experimental investigation.

\section*{Acknowledgements}

We thank Sebastian Diehl for a stimulating discussion. This work was
financially suppported by the FWO Odysseus program.

\end{document}